\documentclass[showpacs,preprintnumbers,amsmath,amssymb,floatfix,prd,11pt]{revtex4}
\usepackage[utf8]{inputenc}
\usepackage{epsfig}
\usepackage{amsmath}
\usepackage[dvipsnames]{xcolor}

\begin{document}
\title{NNLO jet production in neutral and charged current polarized deep inelastic scattering}
\author{Ignacio Borsa}  
\email{iborsa@df.uba.ar}
\affiliation{Universidad de Buenos Aires and IFIBA, Facultad de Ciencias Exactas y  
Naturales, Departamento de Física. Buenos Aires, Argentina.}
\author{Daniel de Florian}  
\email{deflo@unsam.edu.ar}
\author{Iv\'an Pedron}  
\email{ipedron@unsam.edu.ar}
\affiliation{International Center for Advanced Studies (ICAS), ICIFI and ECyT-UNSAM, 25 de Mayo y Francia, (1650) Buenos Aires, Argentina}

\begin{abstract}

We perform the calculation of the fully differential single-jet production in longitudinally polarized deep inelastic scattering (DIS) at the next-to-next-to-leading order (NNLO) accuracy, featuring both neutral current and charged current processes. The computation is done with the projection-to-Born method (P2B), using the next-to-leading order (NLO) dijet calculation and the NNLO DIS structure functions as its main ingredients. We also analyze its phenomenological consequences in the kinematics of the future Electron-Ion Collider (EIC).

\end{abstract}

\maketitle

\section{Introduction}

The advent of the future Electron-Ion-Collider (EIC), set to reach an unprecedented precision in measurements of polarized processes and to extend the kinematical coverage in terms of $x$ and $Q^2$~\cite{Accardi:2012qut}, will provide valuable information for the improvement of our still limited knowledge on polarized parton distributions functions (pPDFs) as well as new insights on the structure of the proton~\cite{Accardi:2012qut, Aschenauer:2012ve, Aschenauer:2015ata, Aschenauer:2020pdk, Boughezal:2018azh}. Particularly relevant amid the questions that the EIC is set to answer is that of the way in which the proton spin is distributed among its constituent, gluons and quarks of different flavors.  
In addition to the purely electromagnetic contributions, for sufficiently large values of the boson virtuality $Q^2$, the electron-proton deep inelastic scattering (DIS) process receives relevant contributions from the exchange of virtual weak bosons $Z$ and $W^{\pm}$. Since the cross sections for neutral and charged current (NC and CC, respectively) DIS involve different quark combinations than their electromagnetic counterpart, data on polarized DIS with electroweak currents can offer complementary information on the proton spin decomposition, allowing to discriminate the individual contributions from quarks and its corresponding antiquarks~\cite{Aschenauer:2013iia}. In that sense, it is worth mentioning the key role that measurements on unpolarized CC DIS already play for the flavor decomposition in modern PDF global analyses, specially for the determination of the strange quark PDF \cite{NNPDF:2021njg, Bailey:2020ooq, Hou:2019efy}. However, there is currently no data available on polarized CC DIS, and most of the flavor separation in pPDFs extractions comes from semi-inclusive DIS (SIDIS) data, which assumes some knowledge on the parton-to-hadron fragmentation functions. Future measurements at the EIC will therefore be a valuable extension of the existing data set for helicity PDFs extractions, providing new constraints for those distributions. In addition, due to the $\gamma Z$-interference with the photon, $Z$ boson exchange in NC processes can be relevant even at lower values of $Q^2$. Thus, NC measurements at the EIC could be used as electroweak precision tests, providing new accurate determinations of the electroweak couplings, as well as act as a probe for the search of beyond-standard-model physics~\cite{Zhao:2016rfu, AbdulKhalek:2021gbh}.

Clearly, these new precision measurements are to be accompanied by accurate theoretical calculations of the corresponding observables. As it is already the standard for Large-Hadron-Collider (LHC) computations, next-to-next-to-leading order (NNLO) calculations are becoming a benchmark for polarized processes, with results already available for inclusive process, such as Drell-Yan~\cite{Ravindran:2003gi} and DIS~\cite{Zijlstra:1993sh, Borsa:2022irn}, the helicity splitting functions~\cite{Vogt:2008yw, Moch:2014sna, Moch:2015usa}, as well as the recent addition of exclusive process like jet production in DIS~\cite{Borsa:2020ulb,Borsa:2020yxh}, $W$ boson production in proton-proton collisions~\cite{Boughezal:2021wjw} and semi-inclusive DIS (in approximate form) \cite{Abele:2021nyo,Abele:2022wuy}. This matching of the precision of the state-of-the-art in polarized cross sections to the one of the corresponding unpolarized counterparts will be crucial, for example, to perform a detailed study of spin asymmetries and eventually reach NNLO accuracy in the extraction of helicity PDFs.

This paper is the latest entry in our series of papers on NNLO fully differential jet production in polarized DIS. Following our previous work, we achieve NNLO results with the Projection-to-Born (P2B) subtraction method \cite{Cacciari:2015jma}, using our recent NLO dijet calculation in NC and CC processes \cite{Borsa:2021afb}, obtained with the extension of the Catani-Seymour dipole subtraction~\cite{Catani:1996vz} to account for polarized initial-state particles, and the NNLO polarized structure functions~\cite{Borsa:2022irn} as key ingredients.

The rest of the paper is organized as follows: in section \ref{sec:higher_order_corrections} we provide the main details on the NNLO calculation and the ingredients needed in the P2B method. In section \ref{sec:single-jets} we present the phenomenological results for NNLO jet production at the EIC in the Laboratory (Lab) frame for both NC and CC processes, analyzing the impact of higher order corrections, its perturbative stability and phenomenological implications. Finally, in section \ref{sec:conclusion} we summarize our work and present our conclusions.

\section{Calculation of higher order corrections}\label{sec:higher_order_corrections}

We specify the DIS process $e(k)+P(p) \rightarrow l(k') + {\rm jet}(p_T,\eta) +X$, where $k$ and $p$ are the momenta of the incoming electron and proton, respectively, and $k'$ is the momentum of the outgoing lepton (either an electron in the NC or a neutrino in the CC case, respectively). We consider both NC and CC processes, with the four-momentum of the exchanged boson given by $q=k-k'$ and its virtuality by $Q^2=-q^2$. The usual inelasticity and Bjorken variables are given by $y=p\cdot q/p\cdot k$ and $x=Q^2/(2 p\cdot q)$. For CC electron-proton scattering it should be noted that, while the kinematics of the outgoing neutrino are not experimentally accessible, the values of $x$ and $Q^2$ can be reconstructed from the hadronic final state using the Jacquet-Blondel method \cite{Aschenauer:2013iia}. The final state jet is characterized by its transverse momentum $p_T$ and its pseudorapidty $\eta$. We analyze the process in the Laboratory frame of the lepton–proton system, where the jet has always non-vanishing transverse momentum with a contribution starting already at ${\cal O}(\alpha_s^0)$. The polarized DIS cross section is given by
\begin{equation}
    d \Delta \sigma \equiv \frac{1}{4} \left( d \sigma^{++} - d \sigma^{+-} - d \sigma^{-+} + d \sigma^{--} \right),
\end{equation}

\noindent where the superscripts denote the helicities of the incoming proton and electron.

The NNLO calculation of single-inclusive jet production in polarized DIS is achieved by combining our fully exclusive computation for dijet production at NLO in polarized $e P$ collisions \cite{Borsa:2021afb} using dipole subtraction \cite{Catani:1996vz} with the corresponding NNLO expression for inclusive polarized DIS (structure functions) by means of the Projection-to-Born \cite{Cacciari:2015jma}. This method subtracts the full matrix element evaluated at the original phase space point but binned in the kinematic corresponding to the Born-projected equivalent for the lowest order process. In the case of DIS, we can calculate the exclusive NNLO single-jet cross section as
\begin{equation}
d\sigma_{\rm jet}^{\rm NNLO} = d\sigma_{\rm 2jet}^{\rm NLO} -d\sigma_{\rm {\rm 2 jet\, P2B}}^{\rm NLO} + 
d\sigma_{\rm jet}^{\rm NNLO , incl}
\label{eq_p2b}
\end{equation}

\noindent where the first term represents the result for the same observable (in this case single-jet production) plus one extra jet at ${\rm NLO}$ accuracy, in which IR divergences are already dealt with through a suitable NLO method (dipoles in this case). The second term is the subtraction corresponding to the same quantity as before but now {\it binned} at the P2B kinematics, which cancels the double-soft and double-collinear divergences. Since the integration of born-projected counterterm is equivalent to the radiative corrections to the inclusive cross section, the last term corresponds to the fully inclusive result at the same desired accuracy. In the case of DIS, that last contribution can be written in terms of the usual structure functions.

The DIS structure functions are defined in terms of the the hadronic tensor
\begin{equation}
\begin{split}
    W^i_{\mu \nu} = & \left( - g_{\mu \nu} + \frac{q_{\mu} q_{\nu}}{q^2} \right) \left[ F_1^i(x,Q^2)- \frac{h}{2} \ g_5^i(x,Q^2) \right] \\
    & + \frac{\left(p_{\mu}-\frac{p \cdot q}{q^2} q_{\mu}\right) \left(p_{\nu}-\frac{p \cdot q}{q^2} q_{\nu}\right)}{p \cdot q} \left[ F_2^i(x,Q^2)- \frac{h}{2} \  g_4^i(x,Q^2) \right] \\
    & - i \epsilon_{\mu \nu \alpha \beta} \frac{q^{\alpha} p^{\beta}}{2 p \cdot q} \left[ F_3^i(x,Q^2)+ h \ g_1^i(x,Q^2) \right],
\end{split}
\end{equation}

\noindent where $h$ is the helicity of the incoming hadron and the superscript $i$ denotes the type of exchanged boson. The $g_2$ and $g_3$ contributions are excluded since they are suppressed by power corrections of $\mathcal{O}(M^2/Q^2)$, with $M$ being the hadron mass, in processes with longitudinal hadron polarization \cite{Forte:2001ph}. Other terms proportional to $M^2/Q^2$ accompanying the $g$'s structure functions were also neglected. In this limit, the unpolarized and polarized inclusive electron-proton cross sections may be written in terms of the corresponding structure functions as \cite{Workman:2022ynf}
\begin{equation}
\begin{split}
    \frac{d^2 \sigma^j}{dx dy} = & \frac{4 \pi \alpha^2}{x y Q^2} \xi^j \left[ Y_+ F_2^j - y^2 F_L^j + x Y_- F_3^j \right], \\
    \frac{d^2 \Delta \sigma^j}{dx dy} = & \frac{4 \pi \alpha^2}{x y Q^2} \xi^j \left[ -Y_+ g_4^j + y^2 g_L^j + 2 x Y_- g_1^j \right],
\end{split}
\label{eq_cross_sections_SF}
\end{equation}

\noindent with $j=\mathrm{NC, CC}$ and $Y_{\pm}=1\pm(1-y)^2$. The coefficients $\xi$ takes the values $\xi^{\mathrm{NC}}=1$ and $\xi^{\mathrm{CC}}=2$, while the NC and CC structure functions correspond to the combinations given in~\cite{Borsa:2022irn}.

While $g_1$ was computed long time ago to NNLO accuracy~\cite{Zijlstra:1993sh}, the parity violating structure functions $g_4$ and $g_5$ were only known with NLO precision~\cite{deFlorian:1994wp,Stratmann:1995fn, Anselmino:1996cd, Forte:2001ph}. However, as a consequence of the axial Ward identity~\cite{Larin:1991tj} they can be obtained from the non-parity violating unpolarized structure functions, as recently presented in~\cite{Borsa:2022irn}. Having the full NNLO knowledge of the polarized parity violating structure functions we can proceed to use them in the P2B subtraction framework.

The calculation is implemented in our code {\tt POLDIS}, which has been extended to include electroweak mediated processes.

\section{Results of Polarized NNLO Single-Jet Production}\label{sec:single-jets}

\subsection{Jet production in neutral current DIS}\label{sec:nc}

In this section we present our results for polarized single-inclusive jet production at NNLO in NC DIS in the Lab frame. We focus primarily on the effects of the inclusion of the $Z$ boson contribution in the NNLO cross section. Following our previous publications, we work in the EIC kinematics, considering electron-proton collisions with beam energies of $E_{e}=18$ GeV and $E_{p}=275$ GeV, and reconstruct the jets with the anti-$k_{T}$ algorithm and $E_T$-scheme recombination ($R=0.8$). The renormalization and factorization scales are fixed at central values of $\mu_{F}^{2}=\mu_{R}^{2}=Q^2\equiv\mu_{0}^2$, with $\alpha_s$ evaluated at NLO accuracy with $\alpha_s(M_z)=0.118$ and using $n_F=4$ active flavors. We require that the jets satisfy the following kinematical cuts:
\begin{equation}
\begin{tabular}{ c }
    $p_{T}>5\ \mathrm{GeV}$,\\
    $|\eta|<3$,
\end{tabular}
\end{equation}

\noindent with $p_T$ and $\eta$ measured in the Laboratory frame, where the LO contribution starts already at $\mathcal{O}(\alpha_S^0)$. The lepton kinematics is restricted by
\begin{equation}
\begin{tabular}{ c }
    $0.04<y<0.95$,\\
    $25\, \mathrm{GeV}^{2}<Q^{2}<1000 \, \mathrm{GeV}^{2} $.
\end{tabular}
\end{equation}

\begin{figure}
 \epsfig{figure= 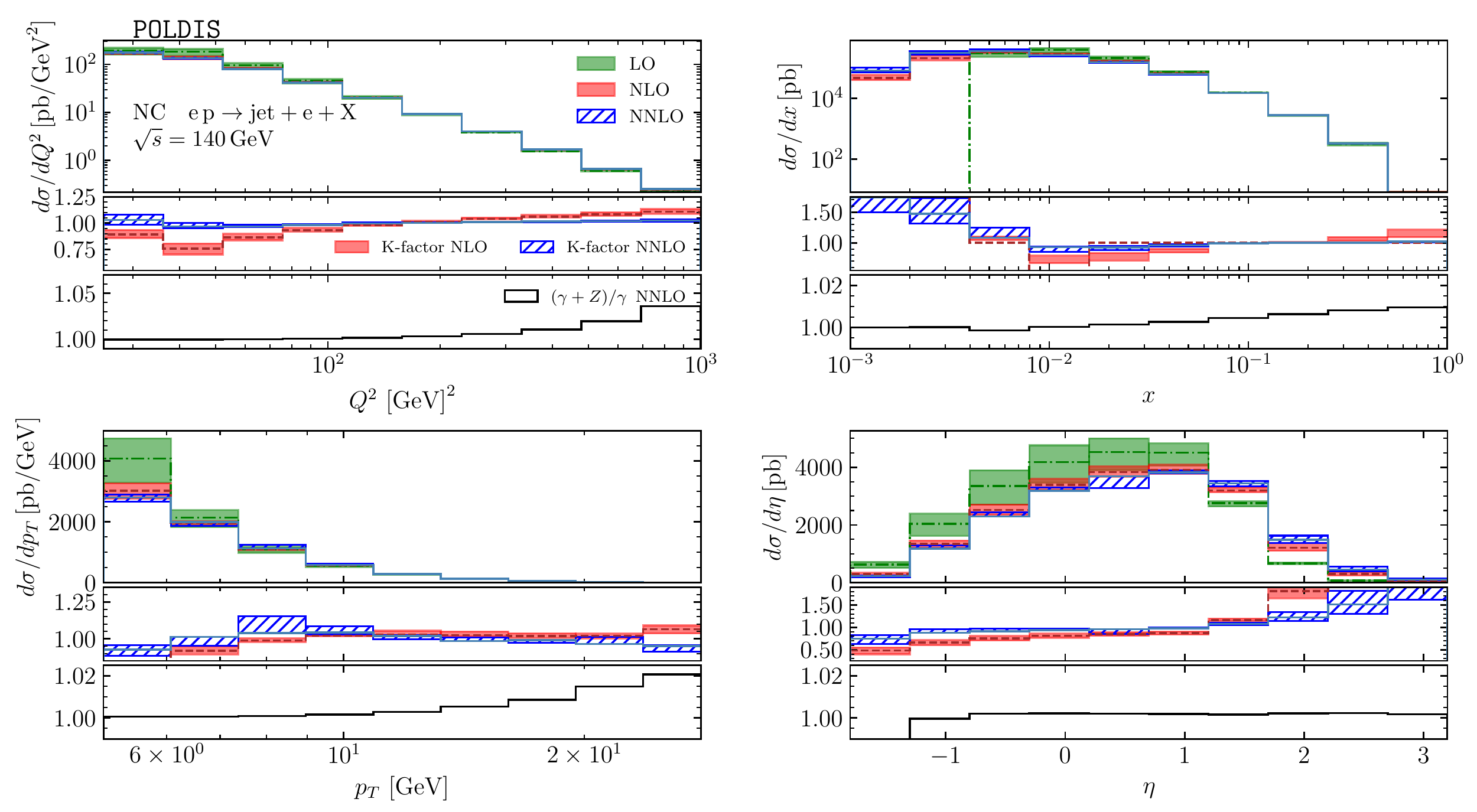, width=0.98\textwidth}
  \caption{Inclusive jet production cross section as distributions in $Q^2$, $x$, $p_T$ and $\eta$, for unpolarized DIS mediated by neutral currents at LO (green), NLO (red) and NNLO (blue). The uncertainty bands represent the theoretical uncertainty of the cross section, obtained by an independent 7-point variation of the renormalization and factorization scales.  The lower boxes show the $K-$factors, i.e, the ratio of each perturbative order to the previous one, as well as the ratio between the NC distributions and those corresponding to pure photon interchange (at NNLO).}\label{fig_dist_nopol}
\end{figure}

\begin{figure}
 \epsfig{figure= 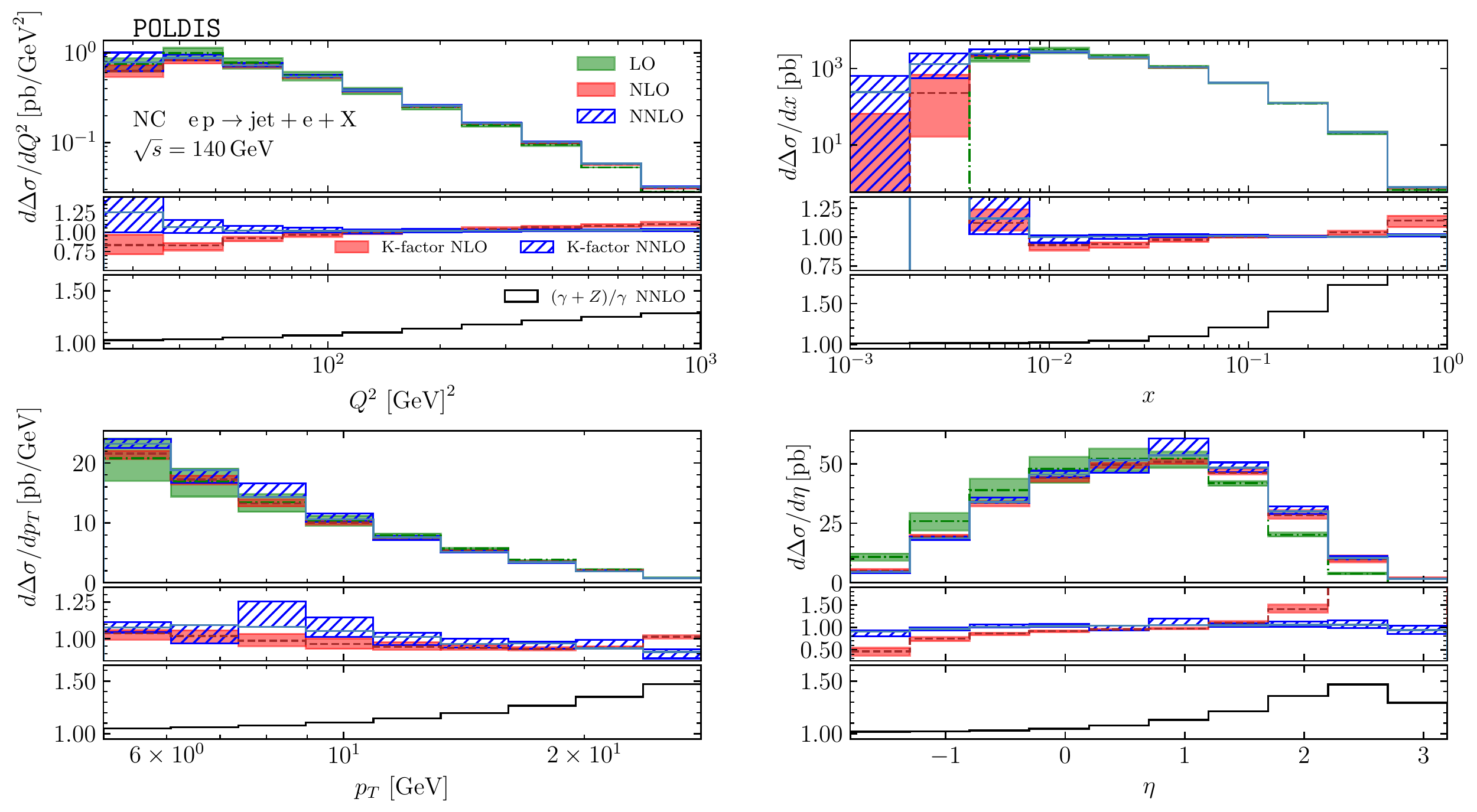, width=0.98\textwidth}
  \caption{Same as Fig. \ref{fig_dist_nopol}, but for the polarized case. }\label{fig_dist_pol}
\end{figure}

For the $Z$ boson we use a mass $M_z = 91.1876$ GeV and a decay-width of $\Gamma_z = 2.4952$ GeV, with an electromagnetic coupling
constant $\alpha = 1/137$ and the Weinberg angle given
by $\sin^2\theta_W = 0.23122$. The parton distributions sets used were the NLOPDF4LHC15 \cite{Butterworth:2015oua} and DSSV \cite{deFlorian:2014yva,deFlorian:2019zkl} for the unpolarized and polarized case, respectively. We note that, since there are no \textit{global} analyses of polarized PDFs available at NNLO, we restrict ourselves to NLO PDFs, in both cases \footnote{While this is not a fully consistent calculation at each perturbative order, it is particularly convenient in order to evaluate the size of the higher order corrections and specially the impact on the asymmetries given the lack of NNLO polarized parton distributions.}.

In Fig.~\ref{fig_dist_nopol} we present the NNLO jet production cross section in unpolarized DIS for processes mediated by the full NC exchange ($Z/\gamma$). They are expressed as distributions in the boson virtuality $Q^2$, the Bjorken $x$, the jet transverse momentum $p_T$, and its pseudorapidity $\eta$. The bands shown correspond to the estimation of the theoretical uncertainty, obtained performing the seven-point variation of the factorization and renormalization scales as $\mu_R,\mu_F = [1/2,2]\, \mu_{0}$ (with the additional constrain $1/2\leq\mu_{F}/\mu_{R}\leq2$). In the lower insets we include the $K-$factors, i.e. the ratio of each successive perturbative order to the previous one ($\sigma_{N^kLO} / \sigma_{N^{k-1}LO}$), and the ratio between the $Z/\gamma$ mediated cross section to the one with pure photon exchange. As it was the case in dijet production, the massive $Z$ boson propagator suppression makes the overall contribution from the $Z$ quite small for the values of $Q^2$ to be covered by the future EIC (reaching the 4\% level only for $Q^2>1000$), and it arises mostly from the $Z\gamma$ interference terms.

Fig. \ref{fig_dist_pol} presents the same distributions as Fig.~\ref{fig_dist_nopol}, but for longitudinally polarized scattering. In terms of the perturbative stability, both the unpolarized and polarized distributions show signs of convergence, with a reduced scale dependence as higher order corrections are considered, and  good overlap between the NNLO cross sections and the NLO ones. The convergence is somewhat spoiled at low $Q^2$ (or, equivalently, low $x$) due to the opening of a new region of phase space starting at NLO, as in the case of photon exchange. This effect is increased in the polarized process due to cancellations between the quark channel and the enhanced gluon contributions at low $Q^2$, leading to higher theoretical uncertainties in that region.

As for the effect of including weak currents, even though the polarized cross section suffers from the same propagator suppression as the unpolarized one, the contribution from the $Z$ boson exchange is much more significant in the former, as can be noted from a comparison of the lowest insets of Figs. \ref{fig_dist_nopol} and \ref{fig_dist_pol}. A similar effect was observed for the dijet production cross sections in \cite{Borsa:2021afb}. In the $Q^2$ distribution, the increase of the cross section ranges between around 5\% at low $Q^2$ and 30\% at high $Q^2$. There are also significant contributions in the high $x$, $p_T$ and $\eta$ regions, even surpassing the 50\% enhancement in some bins. It should be noted that the contributions to the jet cross section stemming from the parity-violating structure functions in the last term of Eq.\ref{eq_p2b}, which are non-zero for $Z$ exchange, are enhanced for polarized DIS. This can be understood from the fact that the polarized cross section involves the difference between the contributions where the initial quark and lepton helicities are parallel and those where they are opposite (this manifests in the $1-(1-y)^2$ factor that weights the $g_1$ structure function in Eq.~\ref{eq_cross_sections_SF}). However, in the parity-violating pieces the opposite is true (hence the $1+(1-y)^2$ factor accompanying $g_4$). For the unpolarized counterpart the argument is reversed, leading to the relative suppression of these parity-violating terms. This effect leads to the overall enhancement of the polarized parity-violating contributions, resulting in a more sizable effect.  

On the other hand, the relevance of the $Z$ boson contribution at moderate and low values of $Q^2$ is, once again, due to the fact that the contributions from different partonic channels have relative signs, leading to cancellations between channels and, therefore, becoming more sensitive to corrections. These channels receive different contributions from the $Z$ boson since, for example, the gluon contribution to parity violating piece of the cross section (related to the $g_4$ and $g_5$ structure functions) cancels after the integration over the phase space, while the quark net contribution is non-zero. This effects result in sizable relative corrections to the polarized cross section due to the inclusion of $Z$ boson exchange.

\begin{figure}[!tb]
 \epsfig{figure= 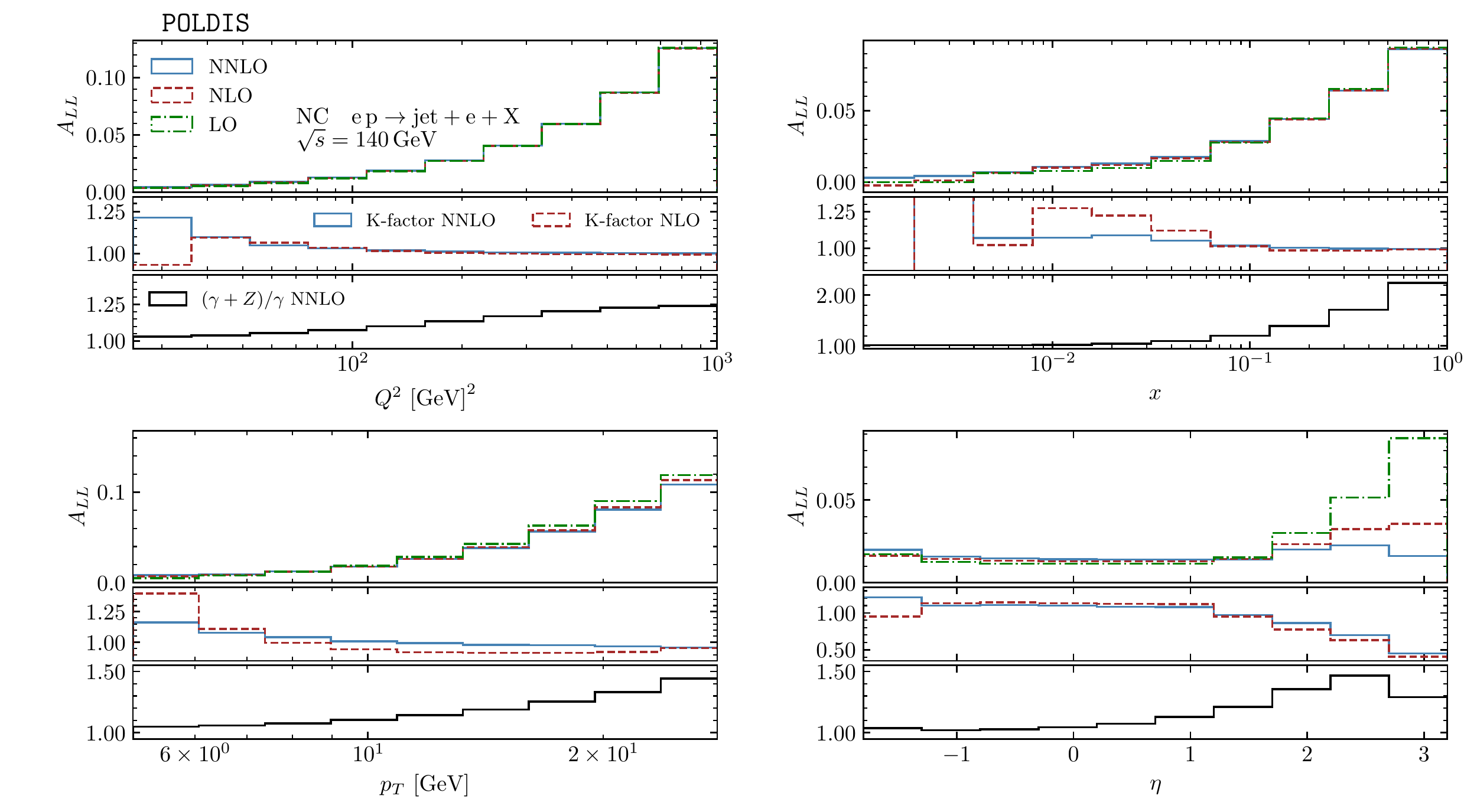, width=0.98\textwidth}
  \caption{Double spin asymmetries for single-jet production, as a function of $Q^2$, $x$, $p_T$ and $\eta$, for DIS mediated by NC. As in the case of Figs.~\ref{fig_dist_nopol}, \ref{fig_dist_pol}, the lower boxes show the $K-$factors as well as the ratio between the asymmetries for full NC  and those corresponding to pure photon exchange.}
  \label{fig_dist_asym}
\end{figure}

The difference in the magnitude of the $Z$ boson contribution in the polarized and unpolarized distributions results in a substantial increase of the double spin asymmetries, presented in Fig.~\ref{fig_dist_asym}, which are defined as the ratio of the polarized to the unpolarized cross sections $A_{LL}=\Delta \sigma / \sigma$. The asymmetry is enhanced at the high $Q^2$ region, where it increases up to a 25\% with respect to the pure photon DIS result, but only receives significant corrections from higher orders at lower $Q^2$. The asymmetries in $x$, $p_T$ and $\eta$ are also enhanced towards higher values, where the contribution from $Z$ in the polarized cross section is more relevant. $K-$factors show a general improvement on the convergence at NNLO, with milder corrections than those of the previous order. The only exception is the significant reduction of the asymmetry in the forward region of high $\eta$, already present in photon exchange, where convergence is spoiled due to the enhancement of soft and collinear gluonic radiation where the resummation of large logarithmic corrections seems to be necessary.

\subsection{Jet production in charged current DIS}\label{sec_cc}

In this section we present our results for polarized jet production at NNLO in CC DIS. The cut imposed on the jets and the reconstructed lepton kinematics are the same as the ones in the previous section. For the $W$ boson we use a mass $M_W = 80.379$ GeV and a decay-width of $\Gamma_W = 2.085$ GeV. The values used for the CKM matrix are $|V_{ud}|=0.9737$, $|V_{us}|=0.2245$, $|V_{ub}|=0.00382$, $|V_{cd}|=0.2210$, $|V_{cs}|=0.987$ and $|V_{cb}|=0.041$.

\begin{figure}[!tb]
 \epsfig{figure= 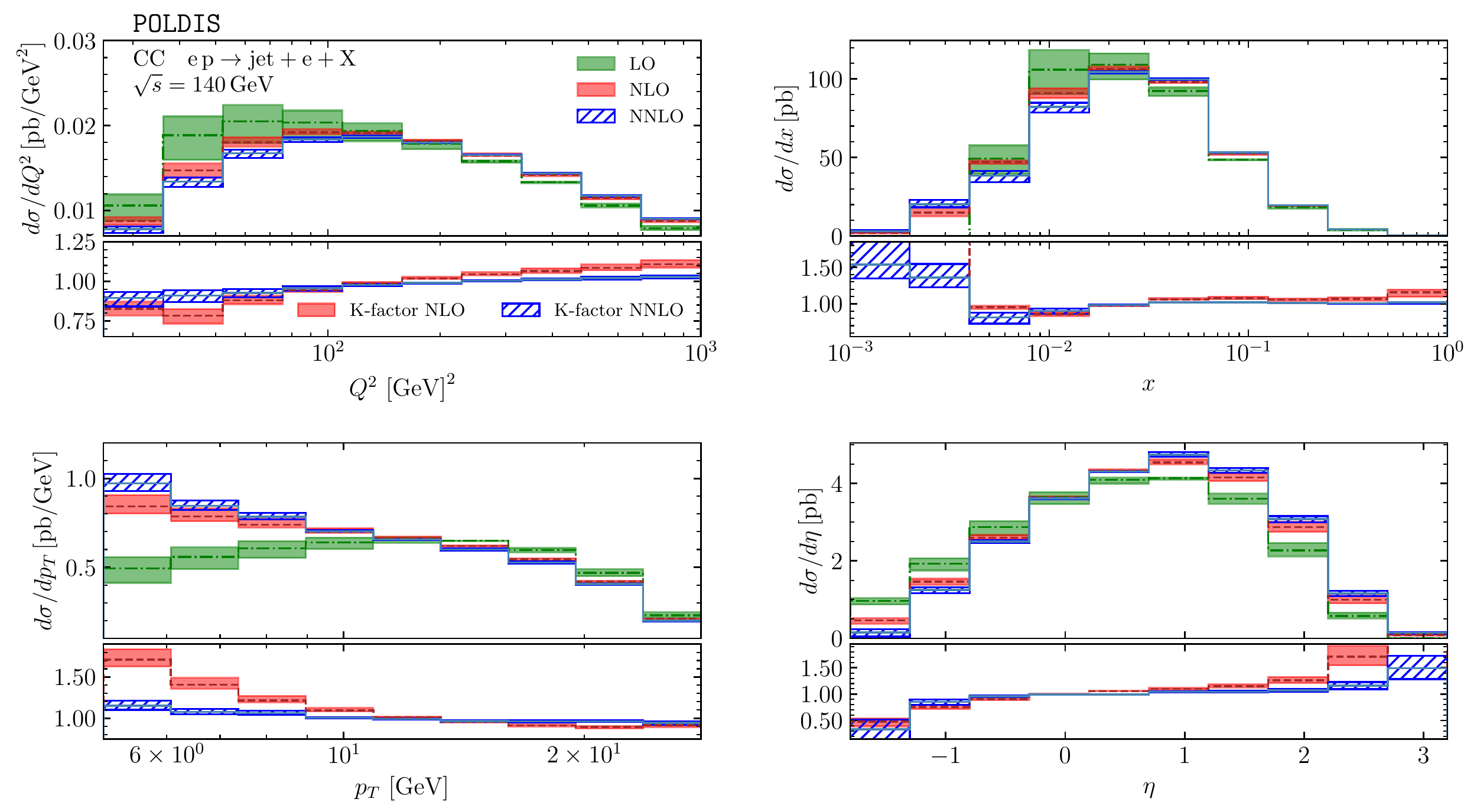, width=0.95\textwidth}
  \caption{Cross section for inclusive jet production in unpolarized, charged current electron-proton DIS, presented as in Fig. \ref{fig_dist_nopol} .} \label{fig_dist_w_nopol}
\end{figure}

We start by studying unpolarized jet production for the typical EIC kinematics. In Fig. \ref{fig_dist_w_nopol} we present the unpolarized cross section as a distribution in the $W$ boson virtuality $Q^2$, Bjorken $x$, the jet transverse momentum $p_{T}$ and its pseudorapidity $\eta$. As in the case of dijet production, the distributions of Fig. \ref{fig_dist_w_nopol} are suppressed compared to the NC ones, particularly at low $Q^2$, due to the mass term in the boson propagator and the lack of interference terms with photon-mediated processes. This also accounts for the shift in the $x$ distributions to higher momentum fractions (since $Q^2$ and $x$ are correlated). As with NC, the distributions show in general a good perturbative convergence, with a reduction in the scale dependence at higher orders. From NNLO, the cross sections at most of the bins overlap with the NLO ones, pointing towards the stabilization of the perturbative series. Once again, that convergence is spoiled at low $Q^2$ as new regions in the phase space become available starting at NLO.

In terms of the $p_T$ and $\eta$ dependence, the effect of higher order corrections is to produce a shift in the distributions towards the forward region, with an increase of the cross section for $\eta \gtrsim 1$ and $p_T \lesssim 10$ GeV. Since at LO the jet transverse momenta is given by $p_T^2 = Q^2\,(1-y)$, the low $Q^2$ suppression results in a pronounced reduction of the cross section at low $p_T$ that is no longer present at higher orders.

\begin{figure}[!tb]
 \epsfig{figure= 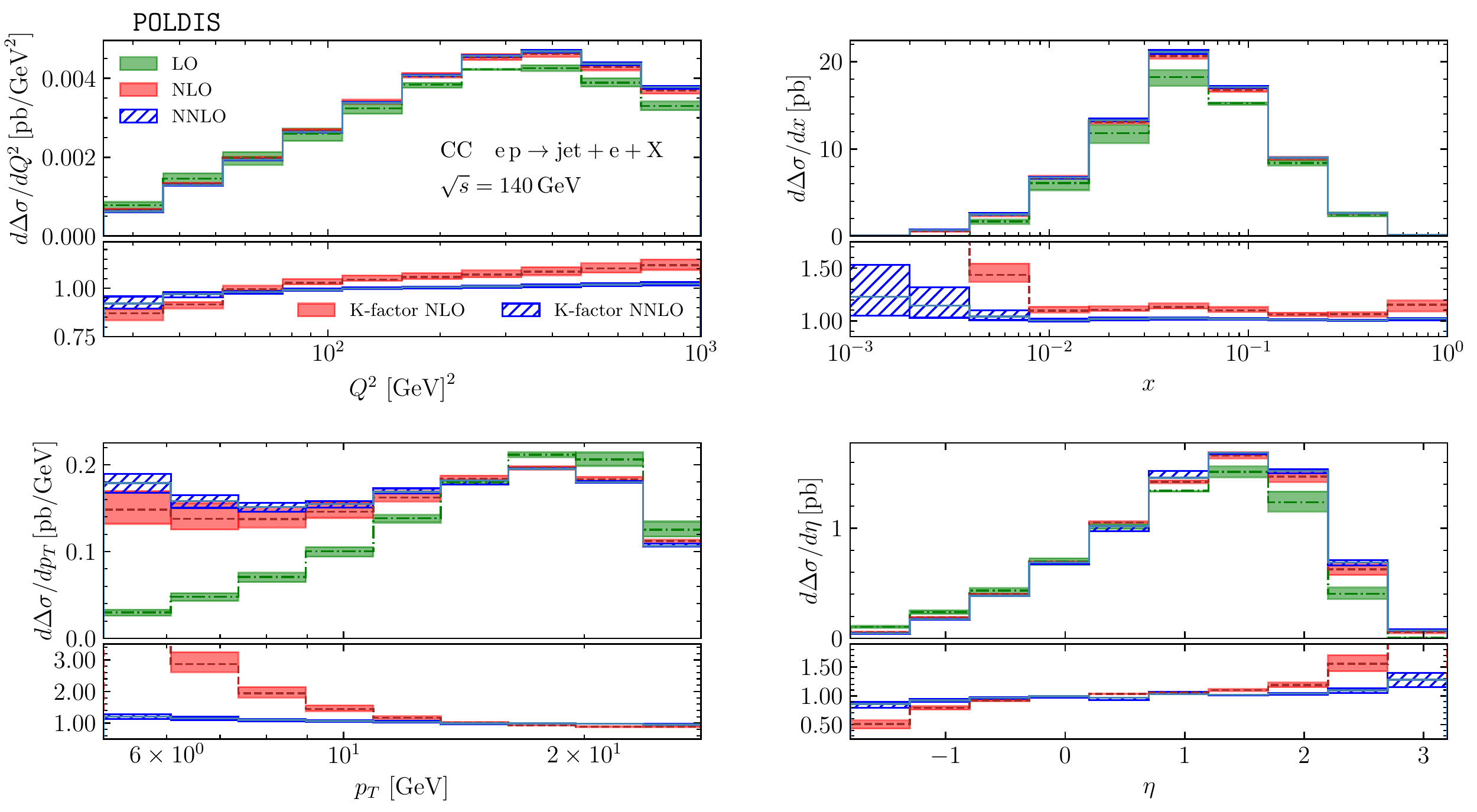, width=0.95\textwidth}
  \caption{Same as in Fig. \ref{fig_dist_w_nopol}, but for polarized electron-proton scattering.}\label{fig_dist_w_pol}
\end{figure}

In Fig.~\ref{fig_dist_w_pol} we present the polarized CC single-jet distributions, in the same fashion as in Fig.~\ref{fig_dist_w_nopol}. In terms of the effect of higher order corrections, the trends are similar to that observed for unpolarized distributions. It is worth noticing that, for CC DIS, the polarized cross section is not as suppressed (compared to the unpolarized one) as in the case of NC or purely photonic DIS. As it was shown in the previous section, the parity violating part of the cross section is enhanced in the polarized case. For CC DIS, this contribution is even more relevant (when compared to its NC counterpart) due to the $W$ boson having equal vector and axial couplings. This milder suppression of the polarized cross section is further amplified by a significant reduction of the cancellation between partonic channels. Since only initial-quark processes contribute to the parity-violating terms, the contributions associated to the gluon channel becomes less relevant.

The main differences between the polarized and unpolarized distributions can be traced back to the different behavior as lower values of $Q^2$ are approached, which is mainly associated to the suppression of the helicity parton distributions at lower momentum fractions. This stronger suppression is ultimately responsible for the low $p_T$ behavior at LO.

\begin{figure}
 \epsfig{figure= 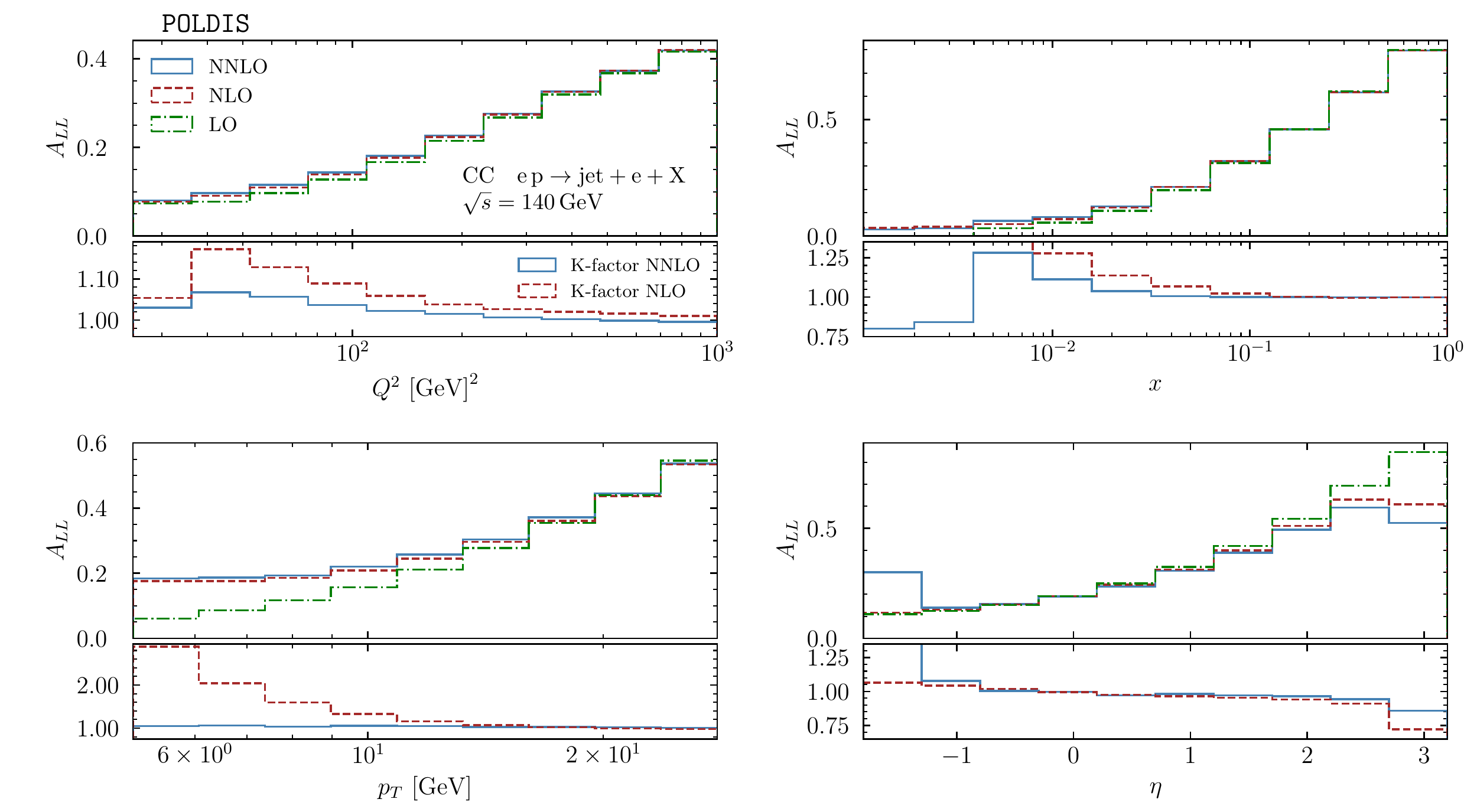, width=0.98\textwidth}
  \caption{Double spin asymmetries as distributions in $Q^2$, $x$, $p_T$ and $\eta$ for CC electron-proton scattering.}\label{fig_dist_asym_w}
\end{figure}

The absence of significant cancellations between the quark and gluon channel in the CC polarized cross section translates into higher values of the double spin asymmetries, which are presented in Fig.~\ref{fig_dist_asym_w}, compared to the NC case. In spite of the small value of the cross section, the asymmetries for charged current DIS are typically greater than $0.3$ for the relevant high-$Q^2$ region, and  can reach values above 50\% at
the end of the spectrum in all the variables studied. This highlight the potential value of CC scattering in the determination of the helicity parton distribution functions, providing additional constraints to them.

\section{Conclusions}\label{sec:conclusion}

In this work, we presented the computation of the fully exclusive single-jet production cross section in (longitudinally) polarized DIS at NNLO, for both neutral and charged current processes. The calculation, implemented in our code {\tt POLDIS}, is done with the projection-to-Born method, combining our NLO polarized dijet computation (achieved by dipole subtraction) with the NNLO inclusive results, expressed in terms of the usual DIS structure functions. The code was used to analyze the phenomenological results in the EIC kinematics. The cross sections were studied as distributions in the virtuality $Q^2$, the Bjorken $x$, the jet transverse momentum $p_T$ and its pseudorapidity $\eta$. 

In terms of the perturbative stability, our calculations for NC DIS show a general overlap between the NNLO and NLO results, as well as a reduction of the scale-dependence for most of explored kinematical region, pointing towards the convergence of the perturbative series. The inclusion of the NNLO corrections mainly leads to a shift of the distributions towards the forward region, enhancing the contributions in the low $p_T$ and high $\eta$ regions. As expected, the contribution of the $Z$ boson exchange, mainly due to its interference with the photon process, only accounts for a small enhancement in the unpolarized cross section, reaching at most 2\% at the highest $Q^2$ bins, where the massive boson exchange is less suppressed. However, due to the nature of the helicity structure on the polarized cross section, parity-violating terms, which only receive contributions from initial-quark processes, are enhanced and mitigate partonic channel cancellations, leading to noticeable contributions even at low $Q^2$. In this case, the enhancement of the cross section due to the $Z$ can reach 30\% for the high-$Q^2$ region, resulting in a sizable increase in the double spin asymmetries.

The results of CC processes also show good perturbative convergence and a general shift of the distribution towards the forward region. The inclusion of higher order corrections is particularly relevant for the correct description of the $p_T$ distribution. Since at LO $p_T$ is correlated to $Q^2$, the mass suppression at low $Q^2$ is translated to a reduction of the LO distribution at low $p_T$, that nonetheless receives large corrections at the following orders. The double spin asymmetries are higher when compared to the NC case, due to the enhanced relevance of the parity-violating terms that do not suffer from channel cancellations between quarks and gluons in the polarized cross section. This results in asymmetries of order 30\%, that can reach results beyond 50\% at higher values of $Q^2$, $x$, $p_T$ and $\eta$.

Data on CC and NC DIS will greatly help to enhance our knowledge on the proton spin structure, providing additional information to disentangle the spin contribution from different flavors of quarks and antiquarks. In that sense, we expect our results to play an important role in future pPDFs determination. Having reached NNLO accuracy for these processes is specially relevant considering  that the cross sections for the most flavor-sensitive experiments in pPDFs global analyses, i.e. semi-inclusive DIS, are currently known up to NNLO only approximately.

The results presented here for both NC and CC DIS highlight not only the relevance of higher order corrections for the precise description of jet production in polarized DIS, but also the potential of electroweak currents to provide an additional probe into the flavor decomposition of the proton spin in the framework of the future precision measurements to be obtained at the EIC.

This work was partially supported by CONICET and ANPCyT.

\bibliography{refs}
\clearpage

\end{document}